\scrollmode

\documentclass{article}
\usepackage{epsfig}
\usepackage{amssymb}

\let\rho\varrho

\begin{document}

\title{Infinite compressibility states in the
Hierarchical Reference Theory of fluids.\\
I.~Analytical considerations}

\author{Albert~Reiner$^{1,2,3}$,\\
${}^1$Institut f\"ur Theoretische Physik and\\
Center for Computational Materials Science,\\
Technische Universit\"at Wien,\\
Wiedner Hauptstra\ss e~8--10, A--1040 Vienna, Austria.\\
${}^2$Teoretisk fysikk, Institutt for fysikk,\\
Norges teknisk-naturvitenskapelige universitet Trondheim,\\
H\o gskoleringen~5, N--7491 Trondheim, Norway.\\
${}^3$e-mail: {\tt areiner@tph.tuwien.ac.at}.}

\maketitle

\begin{abstract}
In its customary formulation for one-component fluids, the
Hierarchical Reference Theory yields a quasilinear partial
differential equation ({\sc pde}) for an auxiliary quantity $f$ that can be
solved even arbitrarily close to the critical point, reproduces
non-trivial scaling laws at the critical singularity, and directly
locates the binodal without the need for a Maxwell construction.  In
the present contribution we present a systematic exploration of the
possible types of behavior of the {\sc pde}\ for thermodynamic states of
diverging isothermal compressibility $\kappa_T$ as the renormalization
group theoretical momentum cutoff approaches zero.  By purely
analytical means we identify three classes of asymptotic solutions
compatible with infinite $\kappa_T$, characterized by uniform or
slowly varying bounds on the curvature of $f$, by monotonicity of the
build-up of diverging $\kappa_T$, and by stiffness of the {\sc pde}\ in
part of its domain, respectively.  These scenarios are analzyed and
discussed with respect to their numerical properties.  A seeming
contradiction between two of these alternatives and an asymptotic
solution derived earlier [Parola {\it et al.{}}, Phys.\ Rev.\ E\ {\bf48},
3321
(1993)] is easily resolved.
\end{abstract}


\section{Introduction}

\label{intro}

Reconciling the vastly different approaches to fluid structure and
thermodynamics afforded by classical integral equation ({\sc ie})
formalisms and renormalization group ({\sc rg}) theory, the Hierarchical
Reference Theory ({\sc hrt},
\cite{b:hrt:1,hrt:1,hrt:2,hrt:3,hrt:4,hrt:8,hrt:10}) presents itself as a
particularly effective instrument
for studying the critical region and liquid-gas phase equilibrium in
simple one-component fluids:
For subcritical temperatures, $T <  T_c$, the usual formulation of the
theory \cite{b:hrt:1,hrt:10} yields density intervals of
rigorously flat free energy and infinite isothermal compressibility
$\kappa_T$ the boundaries of which are readily identified with the
densities $ \rho_v$ and $ \rho_l$ of the coexisting gas and liquid
phases.  The binodal so
found terminates at some temperature $T= T_c$ and density $\rho= \rho_c$
in a liquid-gas critical point characterized by non-classical, partly
Ising-like exponents \cite{hrt:1}.  And far away from the coexistence
region of the phase diagram, {\sc hrt}\ reduces to one of the standard
approximations of liquid state theory, {\it viz.{}}, the popular scheme
commonly known under the names of Lowest-Order $\gamma$ Ordered
Approximation ({\sc loga}, \cite{loga:1,loga:2}) and Optimized Random-Phase
Approximation ({\sc orpa}, \cite{orpa}). {\sc hrt}'s unified treatment of
thermodynamic states so diverse should be contrasted with the
limitations inherent in approaches based on {\sc ie}s\ alone: Close to the
critical point these generally do not have a solution or else develop
major deficiencies, and the binodal is accessible only by way of a
Maxwell construction \cite{b:allg:1}. {\sc rg}\ calculations alone, on the
other hand,
while invaluable for illuminating the scaling relations valid
asymptotically close to the critical point, generally do not allow one
to determine non-universal quantities such as, {\it e.~g.{}}, the {\it
loci} of
the critical point and the binodal.  A theory like {\sc hrt}\ that provides
comprehensive structural and thermodynamic information both close to
and away from the critical point and exhibits a non-trivial scaling
limit is clearly attractive for applications calling for
high-resolution data on the behavior of a fluid in the critical
region.

In the present two-part series of reports we want to have a closer look at
the
solution of the {\sc hrt}\ equations for thermodynamic states of diverging
compressibility, {\it i.~e.{}}, at the critical point and at phase
coexistence. Our motivation for this inquiry is twofold: First of all,
we aim to extend our understanding of the way in which
{\sc hrt}\ achieves its remarkable description of criticality and phase
separation beyond mere invocation of its conceptual ingredients, {\sc rg}\
theory and thermodynamic consistency ({\it v.~i.{}}) in particular.
Instead, it
is on the level of the partial differential equation ({\sc pde}) itself and
the corresponding finite difference ({\sc fd}) approximations used in
practical calculations that we want to understand the mechanism
responsible for the suppression of van der Waals loops and the
emergence of a singular limit of $\kappa_T$ in an extended part of the
phase diagram.  A second reason for our
investigation lies in our earlier work on {\sc hrt}\ and
its numerical side \cite{ar:4,ar:5,ar:th}: For all the merits of the
theory, its practical application has been found to be troubled with
two major difficulties that have been traced to the customary way of
incorporating the core condition of vanishing pair
distribution function for hard core reference systems, and to the
numerical properties of the equations for high compressibility,
respectively. The latter clearly is an issue of prime importance when
focussing on phase separation and the immediate vicinity of the
critical point where $\kappa_T$ diverges, and its severity can only be
assessed on the basis of a thorough understanding of the numerical
process in relation to the properties of the {\sc hrt}\ {\sc pde}.
Evidently,
such an understanding is also highly relevant to the interpretation of
numerical results and the extraction of meaningful and reliable
information from them.

In order to shed some light on these questions, in the present report
we study the analytical properties of the {\sc pde}\ in the presence of a
singular limit of $\kappa_T$.  Relegating some details to the
appendix, after a short introduction to {\sc hrt}\ itself and its
conceptual basis in section~\ref{sec:hrtIntro} we present the {\sc pde},
identify a quantity convenient for following the build-up of infinite
compressibility, and infer the asymptotic scaling relations we base
our work on.  Employing a suitably formalized notion of smoothness, in
sections~\ref{sec:simplistic} through~\ref{sec:stiff} we find a total of
three
scenarios for the gradual build-up of infinite $\kappa_T$, clarify
their respective preconditions, and infer some of their properties
with a view to an eventual implementation by {\sc fd}\ methods.  According
to their most prominent traits we refer to the classes of asymptotic
solutions so found as ``(genuinely) smooth''
(section~\ref{sec:simplistic}), ``monotonous''
(section~\ref{sec:monotonous}),
and ``stiff'' or only ``effectively smooth'' (section~\ref{sec:stiff}),
respectively.  Of these, only the first seems to be compatible with an
earlier analysis of {\sc hrt}'s scaling limit at first sight \cite{hrt:8};
however, a closer investigation into the assumptions implicit in the
simplifications made there leads to a reappraisal of those results
that therefore cannot invalidate either of the remaining two candidate
types of solution (section~\ref{sec:PPR}).

As the considerations outlined above do not take into account the
initial and boundary conditions imposed on the {\sc pde}\ but rather
concern themselves with a summary analysis of the asymptotic behavior
of the various terms in the {\sc pde}\ and of the range of solution types
compatible with these, the all-important question of which of the
scenarios captures the true behavior cannot be answered in the present
contribution.  In part~II\ of our investigation \cite{ar:11} we present a
host of numerical evidence strongly suggestive of a {\sc pde}\
asymptotically turning stiff for thermodynamic states of infinite
compressibility while {\sc fd}\ methods on practical discretization grids
bring about a regularization that leads to an artificially smoothed
solution.  Assertion of stiffness also paves the way for a qualitative
understanding of the relation between specific features of the Fourier
transform of the potential and its numerical properties, clarifies the
special standing of the hard-core Yukawa potential, and leads to a
detailed and self-consistent perception of the process of solving the
equations throughout the domain of the {\sc pde}.  Needless to say, these
findings are highly relevant for the interpretation of numerical
results and the methods of data analysis to be applied to them,
especially when working with non-uniform high-resolution
discretization grids.

\section{Basic relations}

\label{sec:hrtIntro}

As a starting point, let us shortly review the concepts underlying
{\sc hrt}\ when applied to simple one-component fluids, recalling some of
its central notions and establishing the equations we will base our
work on; for a more detailed account of the derivation, its physical
justification and relation to both the {\sc ie}\ formalism and {\sc rg}\
theory
as well as the modifications necessary for dealing with other physical
systems, most notably spin models and fluid mixtures, we refer the
reader to refs.~\cite{b:hrt:1,ar:4} and further references therein. For
consistency with our earlier work on {\sc hrt}\ we employ a number of
notational conventions summarized in the appendix: Most importantly, a
tilde indicates Fourier transformation, and once a symbol has been
introduced we generally omit obvious generic function arguments.  The
appendix also serves as a repository for some of the more cumbersome
analytical expressions as well as for auxiliary definitions and
relations tangential to our reasoning but necessary
to make our presentation self-contained.

Working in the grand-canonical ensemble we consider a system of
particles interacting {\it via}\ pair-wise additive forces taken to derive
from a potential $v(r) = v^{\rm ref}(r) + w(r)$. For the sake of
simplicity,
the potentials are assumed $\rho$ independent, and we restrict the
reference fluid corresponding to $v^{\rm ref}$ alone to a system of hard
spheres of diameter $\sigma$, {\it i.~e.{}}, $v^{\rm ref}(r)$ is infinite
for $r < \sigma$ and
vanishes otherwise. The perturbation $w(r)$ and temperature $T$ enter
the calculation only in the combination $\phi(r)=-\beta\,w(r)$, where
$\beta=1/k_B\,T$ and $k_B$ is Boltzmann's constant.

Based on this splitting of $v$, a momentum space cutoff $Q$ is
introduced by the device of a rather artificial \cite{ar:8} $Q$
dependent potential $v^{(Q)}(r)$ obtained from $v(r)$ by the elimination
from $w$ of all Fourier components $\tilde w(k)$ with $k<Q$, {\it i.~e.{}},
\begin{displaymath}
\begin{array}{c}
v^{(Q)}(r)=v^{\rm ref}(r) + w^{(Q)}(r),
\\
\tilde w^{(Q)}(k) = \left\{\begin{array}{ccc}
\tilde w(k) &:& k>Q \\
0        &:& k<Q.
\end{array}\right.\end{array}
\end{displaymath}
Clearly, the reference and target systems with potentials $v^{\rm ref}$ and
$v$ are obtained in the limits $Q\to\infty$ and $Q\to0$, respectively.
A rather intricate analysis of a resummed perturbation expansion for
the properties of the $(Q - \Delta Q)$ system in terms of those of the
$Q$ system at the same temperature and density in the limit $\Delta Q
\to 0$ finally yields a non-terminating hierarchy of first-order
ordinary differential equations ({\sc ode}s) in $Q$ for the free energy
$A^{(Q)}(\rho)$ and the $n$ particle direct correlation functions
\cite{hrt:2}. Formally, these differential equations allow one to follow
the evolution of structure and thermodynamics of the $Q$ systems when
fluctuations of ever increasing wavelength $1/Q$ are taken into
account, {\it i.~e.{}}, when $Q$ goes from infinity to zero and $v^{(Q)}$
is
transformed from $v^{\rm ref}$ into $v$.

Such an infinite set of coupled {\sc ode}s\ is, of course, hardly tractable
numerically, let alone analytically. As a remedy, a closure on the
two-particle level resembling {\sc loga}/{\sc orpa}, eq.~(\ref{closure}) in
the
appendix, is customarily adopted and combined with only the first
{\sc hrt}\ equation giving the $Q$ dependence of $A^{(Q)}$. As demonstrated
in
ref.~\cite{ar:4}, if {\sc hrt}'s ability to describe phase coexistence is
not
to be lost, it is vital to also incorporate a condition of
thermodynamic consistency into the closure: In {\sc hrt}'s standard
formulation this takes the form of the compressibility sum
rule~(\ref{sumrule}) relating $\kappa_T^{(Q)}$ as obtained by
differentiation
of the free energy to the volume integral of the direct correlation
function at arbitrary $Q$. Due to the density derivatives so
introduced the {\sc ode}s\ at fixed $\rho$ give way to a single {\sc pde}\
in
$Q$ and $\rho$ for the free energy that is to be solved on the
semi-infinite strip ${\cal D}$ where $ \rho_{\rm min} \le \rho \le 
\rho_{\rm max} \wedge
\infty > Q \ge 0$. The precise choice of the initial and boundary
conditions that remain to be imposed at $Q=\infty$, at $\rho= \rho_{\rm
min}$,
and at $\rho= \rho_{\rm max}$ is of no importance for the remainder of this
work, and we refer the reader to refs.~\cite{ar:4,ar:5,ar:th} for a more
detailed discussion of this point.

Discretization of the {\sc pde}\ for the free energy so obtained is
straightforward and yields a computational scheme that can, indeed,
successfully be used for $T> T_c$; for close-to-critical and
subcritical temperatures, however, attempts at a direct solution
invariably fail to produce any results \cite{hrt:4,ar:th}. In order to
remedy this situation, Tau {\it et al.{}}\ \cite{hrt:10} proposed an
alternative
formulation in terms of an auxiliary quantity $f(Q,\rho)$ that is
essentially the first $Q$ derivative of the free energy; just as in
our previous work on {\sc hrt}\ \cite{ar:4,ar:5,ar:th}, in the present
contribution we rely on a slightly different definition for $f$
detailed in the appendix, {\it cf.{}}\ eq.~(\ref{def:f}). Further
specializing to
density-independent potentials and not explicitly including the
core condition that is not expected to be relevant
to the subject of our study, the {\sc pde}\ can be written in quasilinear
form,
\begin{equation} \label{pde:f}
{\partial f\over\partial Q}
=  d_{00}[f;Q,\rho] +  d_{02}[f;Q,\rho]\,{\partial^2f\over\partial\rho^2},
\end{equation}
with initial and boundary
conditions that directly follow from those imposed in the original
formulation.

The rather lengthy expressions for the coefficients $ d_{00}$ and $ d_{02}$
of eq.~(\ref{pde:f}) are to be found in the appendix, as are the defining
relations for a number of auxiliaries.  Among these the quantity
$\varepsilon(Q,\rho) \equiv \bar\varepsilon(Q,\rho) + 1$ is of particular
relevance
to our reasoning: Essentially the exponential of $f$, it turns out
proportional to the isothermal compressibility of the fully
interacting system, {\it cf.{}}\ eq.~(\ref{kappaT}). Infinite
$\kappa_T$ therefore directly implies attendant divergences at $Q=0$
in $\varepsilon$, $\bar\varepsilon$, and $f$, and we have to study the
large-$\bar\varepsilon$ behavior of the {\sc pde}\ if we are to understand
the
description of the critical region afforded by {\sc hrt}\ on the level of
the {\sc pde}.

On the other hand, $f(Q,\rho)$ is guaranteed by the construction of the
{\sc hrt}\
hierarchy to be continuous and finite for every non-vanishing cutoff
$Q$ and to coincide with its limit from above at $Q\to0$ wherever that
limit exists. In other words, for
thermodynamic states $(T,\rho)$ within the coexistence part of the
phase diagram $f$ must take on large finite values for sufficiently
small non-vanishing $Q$, and it must diverge for $Q\to0$: If
$\kappa_T(\rho)$ is infinite, for every threshold $F$ there is a
corresponding cutoff $Q_F(\rho)>0$ such that $f(Q,\rho)>F$ for all
$Q<Q_F(\rho)$, or, less formally,
\begin{equation} \label{f:large:infty}
\kappa_T(\rho)=\infty 
\Longleftrightarrow 
\left\{\begin{array}{lc}
\lim_{Q\to0}f(Q,\rho)=\infty \\
f(Q,\rho)<\infty : Q\sigma>0 \\
f(Q,\rho)\gg1    : Q\sigma\ll1,
\end{array}\right.\end{equation}
which is the very basis of the main arguments of the present report.
Considering large $\bar\varepsilon(Q,\rho)$ and assuming $f$ to diverge
more
strongly than just logarithmically ({\it v.~i.{}},
section~\ref{sec:monotonous}),
inspection of the expressions in the appendix shows both of the
coefficients of the quasilinear {\sc pde}~(\ref{pde:f}) to be of first
order in
$\bar\varepsilon$ whereas $f$ is essentially the logarithm of
$\bar\varepsilon$:
\begin{equation} \label{O:df}
\begin{array}{rl}
d_{02}&{}={\bf O}(\bar\varepsilon),
\\
d_{00}&{}={\bf O}(\bar\varepsilon),
\\
f&{}={\bf O}(1).
\end{array}
\end{equation}
${\bf O}$ and its companion ${\bf o}$ are the usual Landau symbols, and
here as in the remainder of the present series of reports $x={\bf O}(y^a)$
is taken to actually mean
that $a$ is the infimum of all $b$ for which $x={\bf o}(y^b)$ when the
implied limit is $y\to\infty$, or else the supremum when considering
$y\to0$.  Furthermore, qualification of some quantity $x$ as
``essentially'' independent of some other quantity $y$ is equivalent
to the characterization of $x$ as of order ${\bf O}(1)$ in $y$ which does
not rule out a weak, say, logarithmic $y$ dependence of~$x$.  Unless
explicitly stated otherwise, all orders cited are in terms of
$\bar\varepsilon$.

In order to understand the properties of the solution of the {\sc pde}\
around some point $(Q,\rho) \in {\cal D}$ we still need to supplement
eq.~(\ref{O:df}) with an analogous characterization of the behavior of the
remaining term on the right hand side of eq.~(\ref{pde:f}), {\it viz.{}},
${\partial^2 f / \partial\rho^2}$. Lacking any {\it a priori}
information to guide us, in this report we adopt the simple {\it
ansatz}
\begin{equation} \label{O:frr}
{\partial^2f\over\partial\rho^2} = {\bf O}(\bar\varepsilon^r), \qquad r \ge
0,
\end{equation}
a choice that is sufficiently general to admit a consistent
description both of the behavior of the exact solution and of the
computational process yielding a numerical approximation of
it. Inserting eqs.~(\ref{O:df}) and~(\ref{O:frr}) into the {\sc
pde}~(\ref{pde:f}),
comparison of both sides of the equation immediately yields
\begin{equation} \label{O:fq}
\begin{array}{c}
\displaystyle {\partial f\over\partial Q}={\bf O}(\bar\varepsilon^s),
\\
s=\max(1,1+r)=1+r;
\end{array}
\end{equation}
only for $r=0$ is there the added possibility of a cancellation of the
leading terms on the right hand side of eq.~(\ref{pde:f}) that might give
rise to an $s$ less than unity, including $s=r=0$.  At any rate,
neither $r$ nor $s$ may be negative.

Equipped with eqs.~(\ref{O:df}) through (\ref{O:fq}) we are now in a
position
to gain a better understanding of the {\sc pde}'s
workings.  Although the analytical expressions presented and orders
cited below hinge on the assumption (\ref{O:frr}), most of our conclusions
are expected to remain qualitatively valid even in more general
situations, {\it cf.{}}\ section~V\ in part~II\ \cite{ar:11}.

\section{Genuinely smooth solution}

\label{sec:simplistic}

From the {\sc pde}\ itself the $Q$ and $\rho$ scales characteristic of the
variations of $f(Q,\rho)$ in the integration domain ${\cal D}$ are not
apparent:  In particular, we cannot say {\it a priori} whether they
remain bounded from below in an essentially $\bar\varepsilon$ and, hence,
temperature independent way throughout ${\cal D}$  or else scale
like some inverse power at least of $\bar\varepsilon$ and so become
arbitrarily small in part of ${\cal D}$ whenever the temperature falls
below
$ T_c$. Restricting ourselves to
analytical considerations, in this report we will study and present
arguments for both of these types of solution, referring to them as
smooth and non-smooth, respectively.  Clearly, this distinction is
highly relevant to the numerics and therefore of immediate practical
interest: After all, smoothness implies that finite difference ({\sc fd})
schemes are in principle well applicable to the {\sc pde}\ at hand whereas
otherwise local truncation errors will generally be unbounded and at
least an estimate of the global error incurred must be obtained {\it a
posteriori} in order to gauge the significance of any information
extracted from {\sc fd}\ approximations of the {\sc pde}.

\begin{figure}[t]\vbox{\noindent\epsfbox{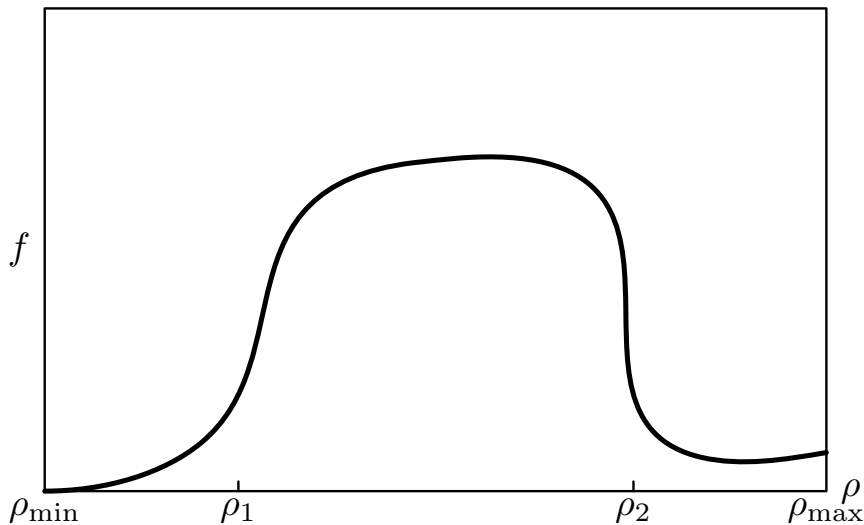}
\bigskip\caption{Sketch of the auxiliary function $f(Q,\rho)$
for fixed and sufficiently small $Q$ in the genuinely smooth
scenario: $ \rho_{\rm min}$ and $ \rho_{\rm max}$ are the densities where
boundary
conditions must be imposed upon the solution of the {\sc pde}; the
density range where $f$ is large extends from $\rho_1$ to
$\rho_2$; and there is a single, rather flat maximum of $f$ for
$\rho_1<\rho<\rho_2$.}\label{fig:sketch}}
\end{figure}

For this section turning to the assumption of $f$ being smooth in the
sense stated --- an assumption that we will repeatedly refer to as
leading to the ``(genuinely) smooth'' scenario ---, we immediately
conclude that $r = s = 0$: If the $Q$ and $\rho$ scales set by $f$ are
bounded from below in the manner indicated, for any linear
differential operator $L$ we can choose $\bar\varepsilon$ independent step
sizes $\Delta Q$ and $\Delta\rho$ such that an {\sc fd}\ approximation of
$L f$ becomes accurate throughout ${\cal D}$. As the estimate so obtained
is
just a linear combination of $f$ values sampled few $\Delta Q$ and
$\Delta \rho$ apart, $L f$ is bound to scale like $f$ and is thus of
order ${\bf O}(1)$. Specializing to $L \equiv \partial^2 / \partial \rho^2$
and to $L \equiv \partial /\partial Q$ we immediately conclude that
$r=0$ and $s = 0$, respectively. --- As a corollary we note that
growth of $f$ in proportion to an inverse power of the cutoff, $f
\propto 1/Q^a$ at fixed $\rho$ with $a>0$, always falls into this
class of solutions.

The main advantage afforded by the presupposition of smoothness is
that it allows us to understand on the level of the {\sc pde}\ the build-up
of infinite $f$ and $\kappa_T$ close to the critical point and in the
coexistence region of the phase diagram: Let us assume that, at some
fixed and sufficiently small cutoff $Q$, $f(Q,\rho)$ is a function of
$\rho$ like that sketched in fig.~\ref{fig:sketch}: continuously
differentiable for all densities, convex and large for densities in a
range with approximate boundaries $\rho_1$ and $\rho_2$ but rather
small elsewhere.  In the region of large $f$ and $\bar\varepsilon$ where
asymptotic reasoning along the
lines of eqs.~(\ref{O:df}) through~(\ref{O:fq}) is applicable, the {\sc
pde}\
coefficient $ d_{00}$ is dominated by terms related to the $Q$ dependence
of the Fourier transforms of the potential and of the reference system
direct correlation function,
\begin{equation} \label{doo:asymptotic}
d_{00} \sim -{\partial(\tilde\phi/\tilde{\cal K})\over\partial Q}\,
{\tilde{\cal K}^2\,\bar\varepsilon^2\over\varepsilon}\,
{\tilde\phi_0^2\over\tilde\phi^4}
\qquad\hbox{for }\tilde\phi\,\varepsilon\gg1,
\end{equation}
{\it cf.{}}\ eq.~(\ref{pde:coeff}).  $ d_{00}$ is therefore expected to be
negative for
most cutoffs less than the position of the first minimum of
$\tilde\phi$, and certainly for very small $Q$: It is by this standard
that the cutoff under consideration must be ``sufficiently small'' as
stated before. --- Trivially, we also see from eq.~(\ref{pde:coeff}) that
$ d_{02}$ is non-positive throughout ${\cal D}$, and negative for
$\bar\varepsilon \ne
0$, as $\tilde\phi_0$ must be positive for the {\sc pde}\ to be stable at
all,
{\it cf.{}}\ section~2.4.1 of ref.~\cite{ar:th} as well as
refs.~\cite{b:hrt:1,hrt:10}.

As for the remaining term on the right hand side of eq.~(\ref{pde:f}),
$\partial^2f/\partial\rho^2$ is negative throughout most of the
density interval of large $f$, and positive on either side as $f$
falls to small values. Furthermore, the maximum of $f$ displayed in
fig.~\ref{fig:sketch} is rather flat so that $\vert \partial^2f /
\partial\rho^2 \vert$ is quite small there. Assuming this curvature so
small that the $ d_{02}$ term in eq.~(\ref{pde:f}) does not reverse the
sign
of the right hand side, we immediately conclude that $\partial
f/\partial Q < 0$ where $f$ is large, corresponding to further growth
of $f$ as $Q=0$ is approached from above.  In this way we can
understand how the {\sc pde}\ implements the gradual build-up of infinite
$\kappa_T$ and the attendant suppression of van der Waals loops in the
free energy.  On the other hand, it becomes clear that this mechanism
depends on
the preservation of the general outline of $f$: In particular, the
stable growth of $f$ is likely to break down once flatness is lost in
the central part of the interval $[\rho_1, \rho_2]$.  A closer look
at the $ d_{02}$ term immediately reveals that it acts to stabilize this
feature of the form of $f$: Taken by itself, $ d_{00}$ strongly favors $f$
to grow most rapidly
close to its maximum. This, however, is prevented by the considerable
increase in the negative density curvature of the solution it would
entail: The correspondingly large contribution $ d_{02} \, \left(
\partial^2 f / \partial \rho^2 \right) > 0$ to $\partial f / \partial
Q$ effectively counteracts $ d_{00}$ and so ensures that $f(Q,\rho)$,
$\rho_1 \lesssim \rho \lesssim \rho_2$, remains flat even as it grows,
just as postulated at the outset.  Close to $\rho_1$ and $\rho_2$,
on the other hand, the {\it r\^ole}\ of the $ d_{02}$ term is quite
different: There the
curvature $\partial^2f / \partial\rho^2$ turns positive so that both
terms on the right hand side of eq.~(\ref{pde:f}) now contribute to the
growth of $f$, thereby rendering the transition to small $f$ ever more
sharply defined as $Q$ progresses towards zero.  By the same token,
intermediate minima in the density range of large $f$ are also
dissolved by the $ d_{02}$ term, which relaxes the precondition of
conformance with fig.~\ref{fig:sketch} somewhat.

This intuitively appearling concept of a stable mechanism of growth of
$f$, as we will henceforth refer to it, so allows us to understand on
the level of the {\sc pde}\ the emergence of infinite $\kappa_T$ within a
clearly demarcated density interval $[ \rho_v,  \rho_l]$.  In
addition, as the coexisting densities $ \rho_v$ and $ \rho_l$ are obtained
as the limits of $\rho_1$ and $\rho_2$ for $Q \to 0$, anything but
continuous inverse compressibility is hard to accomodate within
this picture, which agrees well with the known coincidence of the
{\sc hrt}\ binodal and spinodal for space dimensionality $d=3$
\cite{hrt:8}.
The key {\it r\^ole}\ played by the $ d_{02}$ term in stabilizing the
solution
and locating the densities of the pure phases once more highlights the
importance of the thermodynamic consistency condition~(\ref{sumrule})
underlying the transition from {\sc ode}s\ at fixed $\rho$ to a {\sc pde}\
over
all of ${\cal D}$: Indeed, it should come as no surprise that application
of
an approximation incompatible with the afore-mentioned compressibility
sum rule yields pathological results including negative compressibility
\cite{ar:4}.

Without evidence to the contrary it is then
tempting to subscribe to the seemingly natural assumption of a smooth
solution, $r = s = 0$: Not only does this provide the basis for
understanding the most salient features of {\sc hrt}\ as just discussed, it
is also
what has been found in an early analysis of {\sc hrt}'s scaling limit
\cite{hrt:8}, {\it cf.{}}\ section~\ref{sec:PPR}\ below.  Unfortunately,
however,
this view is
not entirely unproblematic as $r$ and $s$ must both vanish, {\it v.~s.{}}:
As stated in connection with eq.~(\ref{O:fq}), this directly implies
that the right hand side of eq.~(\ref{pde:f}) is affected by massive
cancellation so that the sum of two terms of order ${\bf
O}(\bar\varepsilon)$ each
is reduced to order ${\bf O}(1)$.  Indeed, the relevant signs --- $ d_{02}
\le
0$ throughout ${\cal D}$ [eq.~(\ref{pde:coeff})], $ d_{00} < 0$ for small
$Q$
[eq.~(\ref{doo:asymptotic})], and $\partial^2 f / \partial \rho^2 < 0$
for most densities in $[\rho_1, \rho_2]$ (fig.~\ref{fig:sketch}) --- do
allow such a cancellation.  On the other hand, the rapid growth of the
coefficient
functions $ d_{0i} = {\bf O}(\bar\varepsilon)$ places rather stringent
constraints on the shape of $f$ at constant cutoff: According to
eqs.~(\ref{pde:f}) and~(\ref{O:df}) to~(\ref{O:fq}), $f = {\bf O}(1)$ must
deviate by
terms of order ${\bf O}(1/\bar\varepsilon)$ at most from an exact solution
$\hat
f(Q, \rho)$ of the non-linear {\sc ode}\begin{equation}
\label{cancellation:ode}
{\partial^2 \hat f \over \partial \rho^2}
=
- { d_{00}[\hat f; Q, \rho] \over  d_{02}[\hat f; Q, \rho]}
\end{equation}
for $\rho_1 < \rho < \rho_2$.  Let us now suppose that the initial and
boundary conditions imposed on the {\sc hrt}\ {\sc pde}\ actually lead to
such a
near-solution of the above {\sc ode}\ at some sufficiently small $Q$,
assessment of the likelihood
of which falls outside the scope of this report.  In this case
consistency with $s = 0$ requires the residue $f - \hat f$ of
eq.~(\ref{cancellation:ode}) to vanish as $1/\bar\varepsilon$ when $Q$
further
progresses towards zero, but we have not been able to support this
behavior of the solution on the basis of the explicit
expressions~(\ref{pde:coeff}) for the $ d_{0i}$ and their
properties. Whereas the cancellation necessary for the genuinely
smooth scenario thus certainly cannot be ruled out, it poses far
stricter preconditions than the stable mechanism of growth discussed
before, and both its genesis and stability remain unclear at this
point.

\section{Refined analysis}

\label{sec:refined}

Looking back at the preceding section, we see that the strong point of
genuine smoothness, {\it viz.{}}, the stable mechanism of growth with its
consequences, does not explicitly depend on $s=0$, the very condition
that is the source of the conceptual difficulties just discussed.  It thus
seems pertinent to attempt to salvage these features of the
smooth scenario in a less restricted setting by eliminating the
assumption of vanishing $s$. But rather than separately discussing the
case of $r=0$, $s=1$ we now turn to the full generality afforded by
eqs.~(\ref{O:frr}) and (\ref{O:fq}).  Of course, for
$\partial^2f/\partial\rho^2$ to be of order ${\bf O}(\bar\varepsilon^r)$
with
non-vanishing $r$, $f$ must be a rapidly oscillating function of
$\rho$ and we can only hope to adapt our findings if $f$ resembles
fig.~\ref{fig:sketch} when averaged over oscillations.

Let us consider once more the evolution of $f(Q,\rho)$ within the
region of large $\bar\varepsilon$ as $Q$
progresses towards zero: Combining eq.~(\ref{O:fq}) with the definition
(\ref{def:auxils}) of $\bar\varepsilon$ we find
\begin{displaymath}
\begin{array}{c}
\displaystyle{\partial f\over\partial Q}={\bf O}(\bar\varepsilon^s)={\bf
O}(e^{f s'}),
\\
\displaystyle s'=s\,\tilde u_0^2.
\end{array}
\end{displaymath}
Here the modified exponent $ s'$ takes into account the deviation of
$\tilde u_0(Q)\propto\tilde\phi(Q)$ from unity; for $Q\to0$, $s$ and $ s'$
coincide by virtue of the normalization condition $\tilde u_0(0)=1$. We now
define an auxiliary quantity
\begin{equation} \label{def:d0}
\begin{array}{rl}
d_{0}(Q,\rho)
&{}=\displaystyle
{ d_{00}[f; Q, \rho]
+ d_{02}[f; Q, \rho]\,\left(\partial^2f/\partial\rho^2\right)
\over e^{f s'}}\\
&{}=
{\bf O}(1).
\end{array}
\end{equation}
Existence of $ d_{0}$ is guaranteed, and the {\sc pde}~(\ref{pde:f}) is
fully
equivalent to
\begin{equation} \label{ode:f}
{\partial f\over\partial Q} =  d_{0}\,e^{f s'},
\end{equation}
which no longer involves a derivative with respect to $\rho$.  The signs of
$ d_{0}$ and $\partial f/\partial Q$ always coincide.

If only for the moment we assume $ s'$ constant, eq.~(\ref{ode:f}) can be
formally integrated to yield the solution at all $Q$ given the initial
condition that $f$ be $f_1$ at cutoff $Q_1$, {\it viz.{}},
\begin{equation} \label{f:formal}
f(Q,\rho)
= - {1\over s'}\,
\ln\left(e^{-f_1 s'}+ s'\,\int_Q^{Q_1}\! d_{0}(q,\rho)\,{{\rm d}} q\right),
\end{equation}
which is valid only if $s' \propto s\,\tilde\phi^2$ does not vanish,
{\it i.~e.{}}, if $s > 0$ and $\tilde\phi \ne 0$. The latter condition
effectively
restricts our arguments to cutoffs somewhat below $ Q_{\tilde\phi,1}$, the
smallest of the zeros  $ Q_{\tilde\phi,i}$ ($i=1,2,\ldots$,
$ Q_{\tilde\phi,i} <  Q_{\tilde\phi,i+1}$) of the Fourier transform
$\tilde\phi$ of the
potential; in the limit $Q\to0$ that we are interested in this
certainly poses no problem. With this proviso eq.~(\ref{f:large:infty}) can
be re-cast in form of the relations
\begin{equation} \label{d0:bounds}
\begin{array}{cl}
\displaystyle
-\int_{Q_1}^Q\! d_{0}(q)\,{{\rm d}} q < {e^{-f s'}\over s'}
&\qquad\hbox{for }0<Q_1<Q,
\\\displaystyle
-\int_0^Q\! d_{0}(q)\,{{\rm d}} q = {e^{-f s'}\over s'}
&\qquad\hbox{for }\kappa_T=\infty;
\end{array}
\end{equation}
for the thermodynamic states of interest both
sides of the above inequality are positive if $Q-Q_1$ is comparable to
$Q$.

The remainder of our analytical considerations on the nature of the
{\sc pde}\ for low cutoff in the critical region will be based on
eq.~(\ref{d0:bounds}). As $e^{-f s'}/ s'$ is strictly monotonous in $ s'$,
interval arithmetic is trivial and allows us to tackle in a
straightforward way the problem of the $Q$ dependence of $ s'$: Given the
$\bar\varepsilon$ independent and rather slow variation of
$ s'/s\propto\tilde\phi^2$ as a function of $Q$, for any cutoff interval
considered the range of $ s'$ values is easily found and translated
into an interval of $e^{-f s'}/ s'$. Arguments based upon
eq.~(\ref{d0:bounds}) like those we will present shortly are thus easily
modified to take into account the
non-constancy of $ s'$ simply by applying the least restrictive bound
for any of the $ s'$ values in the $Q$ interval under consideration, an
operation taken for granted throughout the remainder of this series of
reports.

\section{Monotonous growth and logarithmic singularity}

\label{sec:monotonous}

According to its definition (\ref{def:d0}),  the integrand $ d_{0}$ on the
left hand side
of eq.~(\ref{d0:bounds}) is always of order ${\bf O}(1)$. On the other
hand, for $f$ to be a
monotonous function of $Q$ in the asymptotic region, its reduced slope
$ d_{0}$ must always be negative there so that $-\int_{Q_1}^Q\!
d_{0}(q)\,{{\rm d}}
q = {\bf O}(1)\,(Q-Q_1)$. As $Q_1$ is an essentially free parameter to be
chosen from $(0, Q]$, eq.~(\ref{d0:bounds}) shows a quantity scaling like
$Q$ to be bounded from above by another one of order
${\bf O}(\bar\varepsilon^{-s})$, {\it i.~e.{}}, ${\bf O}(Q) < {\bf
O}(\bar\varepsilon^{-s})$. This prompts us
to consider a power law relation between $Q$ and $\bar\varepsilon$, say,
\begin{displaymath}
\begin{array}{c}
Q = {\bf O}(\bar\varepsilon^{-t}),
\\
t \ge s,
\end{array}
\end{displaymath}
corresponding to only a logarithmic
singularity of $f$ at $Q=0$.  In this case, eq.~(\ref{O:df}) no
longer holds due to the cutoff dependence of the prefactors in
eq.~(\ref{pde:coeff}):   To leading order we find
\begin{displaymath}
\begin{array}{rl}
d_{02}&{}={\bf O}(\bar\varepsilon^{1-2t}),
\\
d_{00}&{}={\bf O}(\bar\varepsilon^{1-t}),
\\
f&{}={\bf O}(1)
\end{array}
\end{displaymath}
instead, and the balance equation is modified to
\begin{equation} \label{monotonous:balance}
{\bf O}(\bar\varepsilon^s) = {\bf O}(\bar\varepsilon^{1-t}) + {\bf
O}(\bar\varepsilon^{1-2t})\,{\bf O}(\bar\varepsilon^r).
\end{equation}

If the leading terms on the right hand side of
eq.~(\ref{monotonous:balance}) do not cancel, eq.~(\ref{O:fq}) is thus to
be
replaced by the relation $s = \max(1-t, 1-2t+r)$ with $r \ge 0$ and $t
\ge s \ge 0$. For $r > t$, however, the assumption of monotonous
growth of $f$ leads to inconsistencies: In this case, the right hand
side of the {\sc pde}\ (\ref{pde:f}) is asymptotically dominated by the $
d_{02}$
term. As rapid oscillations in the $\rho$ direction at fixed $Q$
provide the only way for $\partial^2 f / \partial \rho^2$ to become
exponentially large compared to $f$ itself, the second density
derivative of $f$ is bound to be oscillatory in the density range
$\rho_1<\rho<\rho_2$ just as well. Unlike $f$, however,
$\partial^2f/\partial\rho^2$ must change its sign at every swing,
which immediately carries over to $\partial f / \partial Q$ and $ d_{0}$
due to the asymptotic dominance of the $ d_{02}$ term, contrary to the
assumption of $ d_{0}(q)$ being negative for all $q<Q$.

We are thus left with the possibility of $r \le t$, $s = 1-t$;
combining this with the conditions $r \ge 0$, $s \ge 0$, and $t \ge s$
and eliminating $t$ we find the admissible exponent ranges
\begin{displaymath}
\begin{array}{rcl}
0&\le r \le& 1,\\
0&\le s \le& \min(\frac12, 1-r) \le \frac12.
\end{array}
\end{displaymath}

For $r = t$, there is the added possibility of cancellation of the
leading terms on the right hand side of eq.~(\ref{monotonous:balance});
again
eliminating $t$ from the relations $0 \le r = t$, $0 \le s \le 1-t$,
and $t \ge s$ we obtain
\begin{displaymath}
\begin{array}{rcl}
0&\le r \le& 1,\\
0&\le s \le& \min(r, 1-r) \le \frac12.
\end{array}
\end{displaymath}

At any rate, non-vanishing exponent $s$ is compatible with monotonous
growth of $f$ only in the case of a merely logarithmic singularity at
$Q=0$, and in this case there is an upper bound of $\frac12$ for $s$
that holds irrespective of whether the leading terms on the right hand
side of eq.~(\ref{monotonous:balance}) cancel. Furthermore, in this
monotonous growth scenario ${\bf O}(Q) < {\bf O}(\bar\varepsilon^{-s})$,
{\it v.~s.{}}, so that
finally $Q^2\,\bar\varepsilon$ must tend to a finite, possibly vanishing
limit
for $Q \to 0$.

\section{Effective smoothness from stiffness}

\label{sec:stiff}

An alternative is to give up the monotonicity assumption and allow
$ d_{0}$ to alternate in sign: This opens up the possibility of a partial
cancellation of the positive and negative contributions to the
integral in eq.~(\ref{d0:bounds}), and the average of $ d_{0}$ no longer
has
to be of order ${\bf O}(1)$ even though $ d_{0}$ itself still is.  As we
will
now show, this situation implies that the {\sc hrt}\ {\sc pde}\ turns stiff
for
thermodynamic states of infinite compressibility, with $r > 0$ and $s
> 1$. In numerical applications, on the other hand, discretization of the
{\sc pde}\ on
practical grids by necessity induces an artificial smoothing of the
numerical solution that is then characterized by vanishing effective
exponents $ r_{\rm eff}$ and $ s_{\rm eff}$ provided the computation is
able to
reach the limit $Q \to 0$ at~all.

Stiffness of the {\sc pde}\ in that part of ${\cal D}$ where the divergence
of
$\kappa_T$ is built up follows from eq.~(\ref{d0:bounds}): In the scenario
outlined above, the bound $e^{-f s'}/Q s' = {\bf
O}(\bar\varepsilon^{-s})/Q$ on the
mean of $- d_{0}$ over the interval $(0,Q]$ implies that $ d_{0}$ is a
rapidly oscillating function of $Q$ both amplitude and period of which
scale like $1/\bar\varepsilon^sQ = {\bf O}(\bar\varepsilon^{-s})$ [the
possibility of a
non-monotonous logarithmic divergence, $Q \sim \bar\varepsilon^{-t}$ with
$0<t<s$, may be accounted for by replacing $s$ with $s-t > 0$].
Considering eq.~(\ref{ode:f}), for $\bar\varepsilon \gg f \gg 1$ the
solution is
then characterized by oscillations with similar scaling properties
superimposed upon a large, smooth, and most likely monotonously
growing regular part.  These oscillations in the $Q$ direction are
naturally accompanied by oscillations in the $\rho$ direction on
density scales of order ${\bf O}(\bar\varepsilon^{-r/2})$, $r>0$: Vanishing
$r$
would require the evolution of $f$ in $Q$ at densities at fixed
separation to remain synchronous over many, {\it viz.{}}, on the order of
$Q\,\bar\varepsilon^s$ undulations of the function; this seems highly
unlikely, and $r=0$ must be unstable in the setting under discussion.
From eq.~(\ref{O:fq}) we then obtain $r > 0$ and $s = r + 1 > 1$,
where the possibility of a weak $Q$ and $\rho$ dependence of $r$ and
$s$ must also be anticipated.

The scenario just discussed differs from genuine smoothness and
monotonous growth in two important ways: Firstly, the singularity of
$f$ may be stronger than before as there is now no need for the
cancellation
required in section~\ref{sec:simplistic}, nor does boundedness of the mean
of $ d_{0}$ imply boundedness of
the mean slope of $f$ as in
section~\ref{sec:monotonous}. Secondly, non-vanishing exponents $r$ and $s$
mean that the scales
characteristic of the variation of $f$ and, hence, the grid spacings
$\Delta Q$ and $\Delta\rho$ appropriate for the discretization of the
{\sc pde}\ become arbitrarily small as diverging isothermal compressibility
is built up, which obviously has grave repercussions for the
applicability of {\sc fd}\ methods to the {\sc pde}\ at hand.

This last point is worth considering in some detail: Clearly, for the
{\sc fd}\ equations to be a good approximation of the {\sc pde}\ the step
sizes
must go to zero as appropriate inverse powers of $\bar\varepsilon$
determined
by $r$, $s$, and the orders of the local truncation error in $\Delta
Q$ and $\Delta\rho$.  Excluding a very weak divergence of $f$,
however, the rapid growth of $\bar\varepsilon$ certainly renders so fine a
grid impractical \cite{ar:4}, and below some cutoff ${Q_{\Delta Q}}(T,
\rho) > 0$
the numerics can no longer follow the evolution in $Q$ of the true
solution of the {\sc pde}, nor that in $\rho$ below some ${Q_{\Delta
\rho}}(T, \rho) >
0$.  Except right at $\rho_1$ and $\rho_2$, both ${Q_{\Delta x}}$ are
rather
well defined because of the
non-zero exponents $s > r > 0$.  Below the ${Q_{\Delta x}}$, the basic
assumption underlying any {\sc fd}\ method, {\it viz.{}}, applicability of
Taylor
expansions of rather low order, tantamount to smoothness of the
solution on the scales set by the discretization grid spacings, no
longer holds so that a discretization derived on this basis is used
outside its range of validity.  
A chosen implementation of the theory may therefore fail to proceed to
$Q < {Q_{\Delta x}}$ altogether due to
inappropriately large step sizes: In this case no solution is ever
obtained at $Q = 0$ below the critical temperature and neither
criticality nor phase separation can be described, just as reported in
ref.~\cite{hrt:4} and appendix B.1 of ref.~\cite{ar:th}.  Alternatively,
judicious choice of the formulation of the theory and of its
discretization may allow the numerical scheme to solve the {\sc fd}\
equations all the way to vanishing cutoff; it is this very property of
eq.~(\ref{pde:f}) that prompted adoption of the quasilinear form of the
{\sc hrt}\ {\sc pde}\ in the first place, {\it cf.{}}\
section~\ref{sec:hrtIntro}.  Taylor
expansion arguments no longer being applicable, however, {\it a
priori} bounds on the local truncation error are not available, and
the global error may well be substantial.  Furthermore, as any
solution generated in a {\sc fd}\ calculation on a given grid is well
represented on that same grid by definition, we can use the same
arguments as in section~\ref{sec:simplistic} to show that the numerical
results obtained with step sizes of order ${\bf O}(1)$ necessarily
reproduce vanishing {\it effective} exponents $ r_{\rm eff} =  s_{\rm eff}
= 0$.
This reduction of the exponents from $s > r >
0$ for the exact solution corresponds to a smoothing of $f$ that
presumably weakens the singularity, {\it v.~s.{}}, and also suppresses any
oscillations on cutoff and density scales smaller than the step sizes.
Such smoothing is very well known in the numerics of {\sc pde}s\ and in
fact forms the conceptual basis for the highly efficient multi-grid
methods for the solution of integral and differential equations
\cite{ubtuw:181085I,ubtuw:185113I}.

In this ``stiff,'' or only ``effectively smooth'' scenario, any
results obtained by {\sc fd}\ methods then necessarily realize the
genuinely
smooth case of section~\ref{sec:simplistic}, admitting, {\it i.~a.{}}, a
numerical
solution $f \propto 1/Q$.  In this case the artificial smoothing
attendant to the reduction of the exponents directly affects the
solution for $T\le T_c$ at any cutoff lower than ${Q_{\rm smooth}}(T,\rho)
\equiv \max ({Q_{\Delta Q}}, {Q_{\Delta \rho}})$, and the final
results for $\rho_v\le\rho\le\rho_l$.  Given the $ d_{02}$ term of the
{\sc pde}, however, such drastic qualitative ($ r_{\rm eff} \ne r$) and
quantitative ($ s_{\rm eff} \ne s$) changes inside the region of large
$\bar\varepsilon$ must have an effect on the location of the binodal as
well
as on the solution for the pure phases, too. Restriction of large $f$ to
only a rather well-defined density interval $[\rho_1, \rho_2]$, the
{\sc pde}'s parabolic character, and the boundary conditions imposed at
$ \rho_{\rm min}$ and $ \rho_{\rm max}$ nevertheless inspire some hope that
the error
incurred outside the binodal might be limited.  Indeed, in the
presence of stiffness and smoothing this very hope must underlie any
attempt at a numerical solution of the {\sc hrt}\ equations and certainly
has to be justified at least {\it a posteriori} by combining related
calculations and testing the internal consistency of the results
obtained \cite{ar:4,ar:5,ar:th}.

\section{Stiffness and the scaling limit of HRT}

\label{sec:PPR}

So far our considerations have led us to the formulation of three
different possibilities for the asymptotic behavior of the {\sc hrt}\ {\sc
pde}\
for high compressibility states characterized by true smoothness of
the solution, by monotonous growth leading to only a logarithmic
singularity, and by stiffness giving rise to effective smoothness of
$f$, respectively.

At first sight, however, anything but the genuinely smooth solution,
$r = s = 0$, seems to be at variance with the detailed analysis of
{\sc hrt}'s scaling limit in ref.~\cite{hrt:8}: In section~III\,B of that
report we find the prediction that the quantity denoted $u_Q$ there
(corresponding to $-f$) should be quadratic in the density-like
variable $x$ and scale like $Q^{-(d-2)}$ in $d$ dimensions; in our
notation, $f$ should thus grow like $1/Q$ in $d=3$ dimensions,
corresponding to $r = s = 0$, {\it cf.{}}\ section~\ref{sec:simplistic}.

In order to resolve this seeming contradiction let us have a closer
look at the reasoning of ref.~\cite{hrt:8} for $T< T_c$: The pivotal
relation is eq.~(3.9) there, {\it viz.{}},
\begin{equation} \label{3:9}
e^{u_Q}\,{\partial u_Q\over\partial Q} 
= 2\,Q - {1\over2}\,{\partial^2u_Q\over\partial x^2}\,Q^{d-1}.
\end{equation}
For this equation, the direct analogue of our eq.~(\ref{pde:f}), the
authors of ref.~\cite{hrt:8} invoke the validity of ``neglecting
exponentially small terms'' to justify replacing its left hand side by
nought; the asymptotic solution proportional to $Q^{-(d-2)}$ then
follows immediately by separation of the cutoff and density
dependencies.  But this replacement  is legal only if the slope of $u_Q$
actually
remains small in modulus compared to its exponential or, translating
to our notation, if $(1/\bar\varepsilon)\,(\partial f / \partial Q)$ tends
to
zero as $f$ goes to $+\infty$, {\it i.~e.{}}, if $s < 1$ as is the case for
the
solution cited where it actually vanishes. Conversely, eq.~(\ref{3:9}) can
also be read to indicate that its left hand side will be small if and
only if the {\sc pde}'s initial and boundary conditions have caused $u_Q$
at the cutoff considered to almost exactly solve a simple {\sc ode}\ in the
density-like variable $x$, {\it viz.{}},
\begin{displaymath}
{\partial^2u_Q\over\partial x^2} = 4\,Q^{2-d},
\end{displaymath}
which is completely equivalent to the cancellation requirement of
section~\ref{sec:simplistic}, eq.~(\ref{cancellation:ode}) in particular.
The
above directly shows that $\partial^2 u_Q / \partial x^2$ is
proportional to $Q^{2-d} \propto u_Q$ and so of order ${\bf O}(1)$ in
$e^{-u_Q}$; in terms of $f$ this means that $\partial^2f /
\partial\rho^2$ is of order ${\bf O}(1)$ in $\bar\varepsilon$ and, hence,
that $r =
0$.

As the assumptions both of smoothness in the $\rho$ direction, $r =
0$, and of cancellation of the leading terms on the right hand side of
eq.~(\ref{pde:f}), $s < 1$, are thus built right into the derivation of the
scaling solution of ref.~\cite{hrt:8}, conformance of the latter with the
smooth scenario is hardly surprising and certainly does not rule out
any of the other possibilities considered in the present report. As
for the numerics, both genuine ($r=s=0$) and effective
($ r_{\rm eff}= s_{\rm eff}=0$) smoothness suggest an $f$ that
asymptotically grows
like $1/Q$, although in the latter case this does not reflect the
properties of the exact solution of the {\sc pde}.  We are thus led to a
reappraisal of the analysis presented in section~III\,B of
ref.~\cite{hrt:8} as certainly applicable to the {\sc fde}s\ solved
numerically
but not to the true {\sc pde}\ unless $r=s=0$, {\it i.~e.{}}, unless the
cancellation
requirement of section~\ref{sec:simplistic} is also met; with this
proviso the conclusions of ref.~\cite{hrt:8} remain
largely unaffected.  As for the intuitively appealing stable mechanism
of growth responsible for suppression of van der Waals loops, it now
applies to all three scenarios, if only for the numerical process
unless $r=0$; further support for it derives from the $x$-dependence
of $u_Q$ that conforms with the sketch of fig.~\ref{fig:sketch}\ whereas
eq.~(\ref{cancellation:ode}) might well be compatible with a more general
outline.

In view of this resolution of the seeming contradiction between the
earlier analysis of {\sc hrt}'s scaling limit and the possibility of
non-zero exponents $r$ and $s$ none of the three scenarios can be
excluded from further consideration at this point.  On the other hand, the
rather
summary analytical considerations presented here can only identify the
solution types consistent with the asymptotic properties of the {\sc pde}\
for infinite compressibility and explore their preconditions and
consequences, which has been the subject matter of the present report.
The all-important question of which of them is in fact realized in
actual applications of the theory, however, crucially depends on
global aspects of eq.~(\ref{pde:f}) in all of ${\cal D}$ and certainly
cannot be
answered without considering the influence of the initial and boundary
conditions that, after all, uniquely determine $f(Q, \rho)$ throughout
the integration domain.  To arrive at a decision we therefore need to
solve a discretized version of the {\sc pde}, scrutinize the computational
process, and interpret the numerical evidence so gained, a task we
will undertake in part~II\ of our investigation~\cite{ar:11}.

\section*{Acknowledgments}

The author gratefully
acknowledges financial support from {\it Fonds zur F\"or\-der\-ung der
wissen\-schaft\-lichen Forschung} ({\it Austrian Science Fund}, FWF)
under projects~P14371-TPH, P15758-N08, and~J2380-N08.

\appendix
\renewcommand{\theequation}{A\arabic{equation}}\setcounter{equation}{0}

\section{Notational conventions and additional relations}

A most detailed account of the process of re-writing the {\sc pde}\ for the
free energy into the quasilinear eq.~(\ref{pde:f}) can be found, alongside
the explicit expressions for the {\sc pde}\ coefficients $ d_{0i}$, in
appendix~A of ref.~\cite{ar:th}. Further specializing these to take into
account the elimination of the core condition and
the purported density-independence of the potential we obtain
\begin{equation} \label{pde:coeff}
\begin{array}{rl}
d_{00} ={}&\displaystyle
+{\partial\tilde\phi\over\partial Q}\,
\left(
{\tilde\phi_0^2\over\tilde{\cal K}\,\tilde\phi^2}
-{\tilde{\cal
K}\,\bar\varepsilon^2\,\tilde\phi_0^2\over\varepsilon\,\tilde\phi^4}
-{2\,f\over\tilde\phi}
\right)\\&\displaystyle{}
+{\partial\tilde{\cal K}\over\partial Q}\,
\left(
{\bar\varepsilon^2\,\tilde\phi_0^2\over\varepsilon\,\tilde\phi^3}
-{\tilde\phi_0^2\over\tilde{\cal K}^2\,\tilde\phi}
\right)\\&\displaystyle{}
+{\partial^2(1/\tilde{\cal K})\over\partial\rho^2}\,
 
 {Q^2\over4\pi^2}\,{\bar\varepsilon^2\,\tilde\phi_0\over\varepsilon\,\tilde\phi},
\\
d_{02} ={}&\displaystyle{}
- {Q^2\over4\pi^2}\,{\bar\varepsilon^2\over\varepsilon\tilde\phi_0}.
\end{array}
\end{equation}
In the above expressions we have suppressed the obvious function
arguments $Q$ and $\rho$, and we rely on the notational convention
introduced in our earlier work on {\sc hrt}\ according to which
superscripts indicate the system a quantity refers to and a tilde
signals Fourier transformation. We also make use of the following
auxiliary quantities:
\begin{equation} \label{def:auxils}
\begin{array}{rl}
u_0(r) &{}= \phi(r)/\tilde\phi_0,
\\
\tilde\phi_0 &{}= \tilde\phi(0),
\\
\tilde{\cal K}(k,\rho) &{}=\displaystyle{} 
-{1\over\rho} + \tilde c_2^{\rm ref}(k,\rho),
\\
\ln\varepsilon(Q,\rho) &{}= f(Q,\rho)\,\tilde u_0^2(Q)
- \tilde\phi(Q)/\tilde{\cal K}(Q,\rho),
\\
\bar\varepsilon(Q,\rho) &{}=\varepsilon(Q,\rho)-1.
\end{array}
\end{equation}
As far as the definition of ${\cal K}$ is concerned, the ideal gas term
involving $-1/\rho$ is customarily
included in the definition of the hard sphere reference system direct
correlation function $ c_2^{\rm ref}$ throughout much of the literature on
{\sc hrt}. Furthermore, when explicitly taking into account the
core condition ${\cal K}$ is typically augmented by a
truncated series expansion of the deviation of the direct correlation
function of the $Q$ system inside the core from that of the reference
system, a scheme designed to allow one to side-step the need for
costly numerical Fourier transformations that would otherwise arise
from the inversion of the Ornstein-Zernike equation \cite{hrt:4,ar:4}.

The connection to the thermodynamics as well as to the original
formulation of the theory is afforded by the relations linking $f$ and
the auxiliary quantities defined in eq.~(\ref{def:auxils}) to the first
derivative with respect to $Q$ of the free energy $A^{(Q)}$ of the system
at cutoff $Q$: From the expressions given in ref.~\cite{ar:4} one may
easily show that the evolution of $A^{(Q)}$ is given by
\begin{equation} \label{def:f}
\begin{array}{rl}
\displaystyle{
{\partial\over\partial Q}{\beta\,A^{(Q)}\over V}}
&=\displaystyle{
{Q^2\over4\,\pi^2}\,\left(\ln\varepsilon - \tilde\phi\,\rho\right)
\qquad \hbox{for }Q>0,}
\\
\displaystyle{
{\beta\,A^{(0)}\over V}}
&=\displaystyle{
\lim_{Q\to0^+}{\beta\,A^{(Q)}\over V} - {\rho^2\,\tilde\phi_0\over2}.}
\end{array}
\end{equation}
The structure of the $Q$ system follows from the closure
relation that imposes a direct correlation function $ c_2^{(Q)}(r, \rho)$
of
the form
\begin{equation} \label{closure}
c_2^{(Q)} =  c_2^{\rm ref} + \phi^{(Q)} + \gamma^{(Q)}_0\,u_0
\end{equation}
parameterized by the single scalar $\gamma_0^{(Q)}(\rho)$ that is
determined
throughout ${\cal D}$ from thermodynamic consistency in the form of the
compressibility sum rule
\begin{equation} \label{sumrule}
-{1\over\rho} + \tilde c_2^{(Q)}(0,\rho)
=
-{\partial^2\over\partial\rho^2} {\beta A^{(Q)}(\rho)\over V}.
\end{equation}
In the spirit of {\sc loga}/{\sc orpa}, when implementing the core
condition
eq.~(\ref{closure}) is taken to hold only for $r > \sigma$, and the direct
correlation function inside the core is optimized so as to
approximately minimize the pair distribution function there. As shown
in appendix~A.3 of ref.~\cite{ar:th}, the isothermal compressibility
$\kappa_T^{(Q)}$ of the $Q$ system is also readily evaluated and, in the
limit $Q\to0$, found to reduce to
\begin{equation} \label{kappaT}
\kappa_T=\kappa_T^{(0)}
= {\beta\,\bar\varepsilon\over\rho^2\tilde\phi_0}
= -{\bar\varepsilon\over\rho^2\tilde w(0)}.
\end{equation}

\end{document}